# Microscopic Analysis of Homogeneous Electron Gas by Considering Dipole-Dipole Interaction

G. H. Bordbar[1,2] and F. Pouresmaeeli[1]

[1]*Physics Department, Shiraz University, Shiraz* 71454, *Iran**

and

[2] *Department of Physics and Astronomy, University of Waterloo,
200 University Avenue West, Waterloo, Ontario, N2L 3G1, Canada*

**Abstract**

Implying perturbation theory, the impact of the dipole-dipole interaction (DDI) on the thermodynamic properties of a homogeneous electron gas at zero temperature is investigated. Through the second quantization formalism, the analytic expressions for the ground state energy and the DDI energy are obtained. In this article, the DDI energy has similarities with the previous works done by others. We show that its general behavior depends on density and the total angular momentum. Especially, it is found that the DDI energy has a highly state dependent behavior. With the growth of density, the magnitude of DDI energy, which is found to be the summation of all energy contributions of the states with even and odd total angular momenta, grows linearly. It is also found that for the states with even and odd total angular momenta, the DDI energy contributions are corresponding to the positive and negative values, respectively. In particular, an increase of total angular momentum is led to decline in the magnitude of energy contribution. Therefore, the dipole-dipole interaction reveals distinct characteristics in comparison with central-like interactions.

Keywords: dipole-dipole interaction, homogeneous electron gas, second quantization formalism, thermodynamic properties, perturbation theory.

## Introduction

The homogeneous electron gas known as the jellium, has attracted special attention for years. Due to simplicity of the jellium model, it has been widely utilized to describe the valence electrons in metals [1-7]. In this model, electrons interact through the Coulomb interaction in a uniform positive background and the system is made electrically neutral.

Up to now numerous efforts have been done to study the electronic properties of the electron gas. At first attempt, the perturbation theory was employed to compute the ground state energy by Wigner [2]. At high densities, the kinetic energy is more significant in comparison with coulomb potential. Wigner expressed the ground state energy as a function of power series of density. Later, the Hartree-Fock approach turned out to be quite identical. On the other hand, various numerical approaches were introduced to improve the accuracy of calculation for many fermion systems. In order to compute the ground state energy of electron gas, the variational Monte Carlo (VMC) method was employed with a Slater-Jastrow trial function by Ceperley [8]. The diffusion Monte Carlo (DMC) method, which is well known as a more sophisticated approach, was also applied to modify the ground state energy, and numerous thermodynamic quantities were calculated [9]. Further theoretical works were carried out by few researchers to estimate many quantities such as the effective mass [10, 11], the spin susceptibility [11, 12] and the compressibility [13] with variant approaches. Additionally, within the Hartree-Fock approximation, the ground state phase diagram of the electron gas as a function of density was obtained [14]. The ground state properties of a quasi-one-dimensional interacting electron gas, so called nanowire, for both cases with and without uniform magnetic field have been achieved. These studies were performed by Bordbar et al. using the Hartree-Fock and the second quantization methods [15, 16].

---

* Permanent address

In electron gas, the electrons have a spin-spin interaction via dipole-dipole interaction (DDI) with anisotropic and long range characteristics which has taken been into account. The presence of dipolar interaction leads to the emergence of orientation dependence, and mixes the spin and orbital degrees of freedom. Consequently, the DDI, known as non-central interactions, shows more different features in contrast to the central-like interactions [17-19]. As a result of its remarkable properties, the DDI is a good candidate to examine the many-body systems, in particular, electron gas. In several physics literatures, its various impact on the ground state properties of dipolar Fermi gas was theoretically studied [20, 21]. In particular, this interaction induces a deformation of Fermi surface which is stemmed from the exchange energy term in dipolar Fermi gas [20]. In addition, the theoretical studies on the tensor force which has a similar structure to the DDI were carried out in the nucleonic systems. Based on Landau Fermi-liquid theory, the influence of the tensor force on the magnetic susceptibility was also studied in such systems [22, 23].

Despite notable investigations in the field of electron gas, there has been no effort to tackle the current DDI problem. In this paper, the dipole-dipole interaction taken into account in order to examine its effect on the thermodynamic characteristics of interacting electron gas at zero temperature. The anisotropic nature of DDI causes the total angular momentum to be conserved. Due to its particular nature, more complex calculation is required compared to the central-like interactions. We can obtain the ground state energy through the second quantization formalism with the rules of addition of angular momenta. The DDI energy is also calculated separately for each state, and its dependence on the total angular momentum is studied. The findings generally reveal similar behavior in comparison with another theoretical results. Moreover, the influences of density and the total angular momentum on the energy of proposed system are elucidated.

**Method**

A homogeneous unpolarized electron gas with $N$ interacting particles in a uniform positive background with the dipole-dipole interaction is considered.

The Hamiltonian of a neutral electron gas by considering the screening effect is described by

$$H = \sum_{i=1}^{N} \frac{P_i^2}{2m} + \frac{1}{2} e^2 \sum_{i \neq j=1}^{N} \frac{e^{-\mu |r_i - r_j|}}{|r_i - r_j|} + H_b + H_{el-b} + H_{d-d} \tag{1}$$

where the first two terms in equation (1) are electrons Hamiltonians, also $H_{el-b}$ and $H_b$ denoted the interaction of electrons with the background and Hamiltonian backgrounds with particle density, $\rho(x)$, respectively.

$$H_b = \frac{1}{2} e^2 \iint dx dx' \rho(x) \rho(x') \frac{e^{-\mu |x - x'|}}{|x - x'|} \tag{2}$$

$$H_{el-b} = -e^2 \sum_{i=1}^{N} \int dx \, \rho(x) \frac{e^{-\mu |x - r_i|}}{|x - r_i|} \tag{3}$$

To ignore the surface effects, we consider the thermodynamic limits for the system $(L \to \infty, N \to \infty)$. In this condition, the convergence coefficient, $\mu$, approaches zero.

In a homogeneous electron gas, the dipole-dipole interaction (DDI) is occurred between magnetic dipole moment of electrons, $d = d_0 S$, where $S$ is the spin operator and $d_0$ is the strength of magnetic dipole moment. For two dipole moments $d_i$ and $d_j$, the long range and anisotropic dipolar interaction is then given by

$$H_{d-d} = \frac{1}{2} \frac{\mu_0}{4\pi} \sum_{i \neq j=1}^{N} \frac{d_i \cdot d_j - 3(d_i \cdot \hat{r})(d_j \cdot \hat{r})}{|r_i - r_j|^3} \tag{4}$$

In the above equation, $\hat{r} = \frac{r}{|r|}$, where $r = r_i - r_j$ is relative position of the electrons.

At high densities, the magnitude of kinetic energy is larger than long range interactions. The energy of the many body system can be calculated via the perturbation theory, where the Hamiltonian is presented in second quantization formalism by terms of the creation and annihilation operators.

For this system, each quantum state is represented by $\beta \equiv k, s, m_s$. So, the representation of $H_{d-d}$ in second quantization form is [24]

$$H_{d-d} = \frac{1}{2} \sum_{\beta_1 \beta_2 \beta_3 \beta_4} a^\dagger_{\beta_1} a^\dagger_{\beta_2} \langle \beta_1, \beta_2 | V | \beta_3, \beta_4 \rangle a_{\beta_4} a_{\beta_3} \tag{5}$$

Here, $V$ indicates the two-body interaction and is defined as follows

$$V = C_d \frac{\hat{S}_{12}}{|r_1 - r_2|^3}, \tag{6}$$

where

$$C_d = \frac{\mu_0 d_0^2}{4\pi}, \qquad \hat{S}_{12} = (S_1 \cdot S_2) - 3(S_1 \cdot \hat{r})(S_2 \cdot \hat{r}) \qquad (7)$$

Noticed that $C_d = 0.54 \times 10^{-4}\, evA^3$ is the strength of interaction. It is evident from above equations that the matrix elements with respect to single particle wave function are required.

For a 3D electron gas, the single particle wave function consists of a plane wave function and spin wave function, $\chi_{m_s}$, in the volume of $\Omega$.

$$\psi_{k\, m_s}(r) = \langle r | k, s\, m_s \rangle = \frac{1}{\sqrt{\Omega}} e^{ik \cdot r} \chi_{m_s} \qquad (8)$$

It is noticed that in thermodynamic limit, $H_{el-b}$ and $H_b$ have no contributions in energy of the system.

Considering the orientation dependence of $\hat{S}_{12}$ operator (the dipole-dipole interaction) and conservation of the total angular momentum, a different method from the central-like interactions should be employed. As a result, a plane wave function can be expanded in terms of eigenstates of the total angular momentum operator according to the rules for addition of angular momenta.

$$e^{ik \cdot r} = 4\pi \sum_{L=0}^{\infty} \sum_{m_L=-L}^{L} i^L j_L(kr) Y_L^{m_L *}(\hat{\Omega}_k) Y_L^{m_L}(\theta,\phi). \qquad (9)$$

By employing the identity operator in the ket space of $J$, $L$ and $S$, and converting the leading term into the center of mass system, the resulting expression is more simplified.

To perform the calculations, we consider the following changes in the variables:

$$K = k_1 + k_2, \qquad K' = k_3 + k_4, \qquad \kappa = \frac{k_1 - k_2}{2}, \qquad \kappa' = \frac{k_3 - k_4}{2}. \qquad (10)$$

The matrix elements in equation (5) are straightforwardly found as

$$ME = \langle k_1 s_1 m_{s_1}, k_2 s_2 m_{s_2} | V | k_3 s_3 m_{s_3}, k_4 s_4 m_{s_4} \rangle$$
$$= \frac{C_d (4\pi)^2}{\Omega} \sum_{SM_s} F_{SM_s}(\kappa, \kappa') \{C_{SM_s}^{s_1 m_{s_1} s_2 m_{s_2}} C_{SM_s}^{s_3 m_{s_3} s_4 m_{s_4}}\} \delta_{KK'} \qquad (11)$$

where

$$F_{SM_s}(\kappa, \kappa') = \sum_{Jm_J} \sum_{\substack{Lm_L \\ L'm_{L'}}} V_{L,L'}^{Jm_J S} \{C_{Jm_J}^{Lm_L SM_s} C_{Jm_J}^{L'm_{L'} SM_s}\}$$
$$Y_L^{m_L}(\hat{\Omega}_\kappa) Y_{L'}^{m_{L'} *}(\hat{\Omega}_{\kappa'}) \int i^{L-L'} j_L(\kappa r) \frac{1}{r^3} j_{L'}(\kappa' r) r^2 dr. \qquad (12)$$

Here $j_L(\kappa r)$ is the Bessel function of order $L$ with wave number $\kappa$, $Y_L^{m_L}(\hat{\Omega}_\kappa)$ is the spherical harmonic wave function and $V_{L,L'}^{Jm_J S} = \langle Jm_J, LS | \hat{S}_{12} | Jm_J, L'S \rangle$ are the matrix elements of $\hat{S}_{12}$ which vanish for the singlet state. Also $C_{Jm_J}^{Lm_L SM_s}$ and $C_{SM_s}^{s_1 m_{s_1} s_2 m_{s_2}}$ are respectively the Clebsch-Gordan coefficients concerning the addition of $L$, $S$ and $s_1$, $s_2$.

At high densities, the perturbation theory is applied to the calculation of the first order ground state energy which may be determined by evaluating the expectation value of Hamiltonian,

$$E^{(2)} = \langle F|\hat{H}_{d-d}|F\rangle \tag{13}$$

where $|F\rangle$ is the normalized ground state which is characterized by quantum numbers of system.

Now, the expectation value of the number operator is used to determine the Fermi momentum of particles,

$$N = \sum_{k,m_s} \theta(k - k_f) = \frac{\Omega k_f^3}{3\pi^2} \tag{14}$$

where $\theta(x)$ is the step function.

Consequently, the first order perturbation energy of DDI leads to

$$E^{(2)} = \langle F|\hat{H}_{d-d}|F\rangle = \frac{1}{2}\sum_{kpq}\sum_{\substack{m_{s_1}m_{s_2}\\m_{s_3}m_{s_4}}} ME \langle F|a^\dagger_{k+q,m_{s_1}} a^\dagger_{p-q,m_{s_2}} a_{p,m_{s_4}} a_{k,m_{s_3}}|F\rangle \tag{15}$$

By changing the variables,

$$\kappa' = \frac{k-p}{2}, \qquad \kappa = \kappa' + q. \tag{16}$$

Since the momentum of each particle should be less than Fermi momentum, we have the following conditions for this system,

$$\begin{aligned}&1) k+q = p, \quad m_{s_1} = m_{s_4}, \quad m_{s_2} = m_{s_3} \\ &2) k+q = k, \quad m_{s_1} = m_{s_3}, \quad m_{s_2} = m_{s_4}\end{aligned} \tag{17}$$

Under the above conditions, the energy of system reaches the nonzero value.

Then we have

$$\begin{aligned}&\langle F|a^\dagger_{k+q,m_{s_1}} a^\dagger_{p-q,m_{s_2}} a_{p,m_{s_1}} a_{k,m_{s_2}}|F\rangle \\ &= -\langle F|\hat{n}_{k+q,m_{s_1}} \hat{n}_{k,m_{s_2}}|F\rangle \\ &= -\theta(k_f - |k+q|)\theta(k_f - k)\end{aligned} \tag{18}$$

By employing the orthogonality and symmetry relations of Clebsch-Gordan coefficients, one can obtain the explicit expression for equation (15),

$$E^{(2)} = -\frac{C_d \Omega}{6\pi^3} \sum_{J=0}^{\infty} \sum_L \sum_{S=0,1} (-1)^L (2J+1) k_f^3 V_{L,L}^{Jm_JS}$$

$$\int_0^{2k_f} q^2 (1 - \frac{3q}{4k_f} + \frac{q^3}{16k_f^3}) dq \int_0^\infty j_L(\frac{q}{2}r)^2 \frac{1}{r} dr. \tag{19}$$

It is necessary to note that for propagation of the plane wave in the z direction, the condition $m_L = 0$ is satisfied. In the preceding equation, the complicated integral can be solved by analytical methods with Mathematica.

According to the law of addition of angular momenta for a given value of $J$ in the triplet spin state, the orbital angular momentum $L$ can only take three values, $J-1$, $J$ and $J+1$. The diagonal elements of $\hat{S}_{12}$ can be calculated as follows [25],

$$V^{Jm_J 1}_{J+1,J+1} = \frac{2(J+2)}{(2J+1)}, \qquad V^{Jm_J 1}_{J,J} = -2, \qquad V^{Jm_J 1}_{J-1,J-1} = \frac{2(J-1)}{(2J+1)}. \qquad (20)$$

It is obvious that the dipolar interaction has no effect on the spin singlet state. Substituting Eqs. (14) and (20) into Eq.(19), the DDI energy can be derived as,

$$\begin{aligned}E^{(2)} &= \sum_{J=0}^{\infty} E_J^{(2)} = -\frac{C_d \Omega \pi}{4} \sum_{J=0}^{\infty} \sum_{L=J-1}^{J+1} \frac{(-1)^L (2J+1)}{L(L+1)} V^{Jm_J 1}_{L,L} \rho^2 \\ &= \frac{C_d \Omega \pi}{4} \{2 + \sum_{J=1}^{\infty} (-1)^J \frac{4(2J+1)}{J(J+1)}\} \rho^2\end{aligned} \qquad (21)$$

where $\rho$ is the total number density and $E_J^{(2)}$ is called the DDI energy contribution in each state corresponding to the energy of dipole-dipole interaction in the states with the various total angular momentum. The first term in the second line of Eq. (21) is the DDI energy for state $J=0$.

It is revealed that all states with even and odd total angular momenta contribute to the DDI energy. The magnitude of DDI energy in each state depends on the value of the total angular momentum. In addition, the final summation term in Eq. (21) converges to a constant value at higher value of $J$. Eq. (21) reveals that the DDI energy of electron gas is proportional to the volume of container, to the square of density and to the square of magnetic dipole moment. Our findings are in agreement with the DDI energy in dipolar Fermi gas [20, 26]. It should be noted that the DDI energy per particle, $E^{(2)}/N$, is linearly proportional to density.

Therefore, the total ground state energy per particle of the interacting electron gas with the dipole-dipole interaction is readily obtained.

$$\begin{aligned}\frac{E}{N} &= \frac{E^{(0)}}{N} + \frac{E^{(1)}}{N} + \frac{E^{(2)}}{N} \\ &= \frac{3}{10m}(3\pi^2 \rho)^{2/3} - \frac{3e^2}{4\pi}(3\pi^2 \rho)^{1/3} - \frac{C_d \pi}{2}\rho\end{aligned} \qquad (22)$$

**Results and Discussion**

To compute the thermodynamic properties of an interacting electron gas including the dipole-dipole interaction, the ground state energy is calculated. The results at different densities are presented in this section. The variation of kinetic energy per particle versus density is illustrated in Fig. 1. The results show that the kinetic energy monotonically increases by increasing density.

Due to important role of the exchange energy in bounding of the system, the exchange energy per particle as a function of density as shown in Fig. 2. It is seen that as density increases, the magnitude of exchange energy increases. This point stems from the appearance of negative spatial correlation which arises from the repulsive interaction among the fermionic particles. For density value about $\rho = 0.064 A^{-3}$, the kinetic energy and the exchange energy reach $E^{(0)} = 3.50059 ev$ and $E^{(1)} = -4.254\, ev$, respectively. It is clear that the contribution of exchange energy is significant in comparison with the contribution of kinetic energy. Accordingly, considering the recent term in the total energy of system is unavoidable.

The variation of dipole-dipole energy per particle in the triplet spin state as a function of density is plotted in Fig. 3. It can be seen that with increase of density, the magnitude of DDI energy grows linearly. This behavior agrees with another DDI energy results in dipolar Fermi gas at zero temperature [20, 21, 26]. In the range of metallic densities, the DDI energy reveals the diminutive impact against other interaction energies. Consequently, to achieve more accurate computations of the ground state energy of the interacting electron gas, utilizing the aforementioned interaction is beneficial. It is noted that the matrix elements of $\hat{S}_{12}$ in the singlet spin state reach zero. In the singlet state, no orientation with any priority has been observed. It implies that the triplet spin state has the main contribution in the dipole-dipole interaction.

In Fig. 4, the contribution of DDI energy per particle for various states, $J = 0$ to $J = 10$ versus density is plotted. The results indicate that with enhancement of density, the contribution of energy in each state increases, where the DDI energy contribution has the positive and negative values corresponding to the even and odd total angular momenta, respectively. This phenomenon has been also observed for tensor force in two-nucleon system formerly [17-19]. Due to the invariance of the DDI against the spatial inversion, the parity is conserved, the spatial wave functions with even or odd parity confirm this finding. The states with even (odd) angular momentum has even (odd) parity. It is necessary to note that the factor $(-1)^J$ in Eq. (21) indicates the parity of system, and consequently the DDI energy is found to be a mixture of all partial energies with even and odd total angular momenta.

As can be seen in Fig. 5, the Dipole-Dipole interaction acts as a repulsive interaction for the spins in a side by side configuration, possible in states with odd values of $J$. However, the DDI is an attractive potential if spins are parallel to their relative distance and the system is in state with even values of $J$. Hence, distinct behavior emerged by the dipole-dipole interaction in comparison with the central potentials.

The influence of the total angular momentum on the partial DDI energy per particle in the equilibrium density $\rho = 0.014 A^{-3}$ is displayed in Fig. 6. For states with even (odd) values of $J$, this quantity has positive (negative) values which can be also observed in Fig. 4 and can be found in some previous works [17-19]. It is evident that the magnitude of partial DDI energy in odd $J$ states achieves the relatively smaller values comparing to those with even $J$ states. The energy differences are due to the matrix elements of $\hat{S}_{12}$ arising from overlapping of states. Moreover, an increase of the total angular momentum induces reduction of the magnitude of partial energy except in the case of $J = 0$ with allowed value of $L = 1$, and the corresponding curve converges to a certain value. Based on the conservation law of the total angular momentum and assuming $\Delta m_L = 0$, the spin angular momentum remains unchanged (and no spin flip).

In previous articles, the energy of system has been discussed macroscopically, and no dependence on the total angular momentum has been considered. As it is seen in Fig. 4 and Fig. 6, the energy of many-body system is studied by microscopic analysis. To be able to investigate the contribution of partial DDI energy in various states separately, the microscopic analysis is applied. The mentioned analysis can be used not only for dipolar interactions but also for a wide variety of many-particle interactions.

The density dependence of the ground state total energy per particle of the interacting electron gas is shown in Fig. 7. At low densities, the exchange energy plays a dominant role, and this low dense structure behaves as an ideal gas. Conversely, the kinetic energy has more dominant effect compared to others in high density. The ground state energy varies from negative values to positive values. It is obvious that the minimum point in the energy curve which correspond to the equilibrium state of system occurs at $\rho = 0.014 A^{-3}$. At this point, the value of minimum energy are sufficiently close to the experimental data for sodium [27].

**Summary and Conclusion**

To study the role of the dipole-dipole interaction (DDI) in the thermodynamic properties of interacting electron gas, the perturbation method was utilized. Due to structural form of the DDI, different procedures are used by the well-known law for addition of angular momenta. The explicit expressions were obtained for the ground state energy and the DDI energy based on the second quantization formalism. The contribution of DDI energy in each state (partial energy) was also determined by microscopic analysis. The DDI energy was represented as a sum of partial energies with even and odd total angular momentum. The findings of our research are quite convincing in comparison with those reported in the literatures [17, 20, 26]. The effects of two agents on the DDI energy were examined. The results indicated that the magnitude of DDI energy increases linearly as the density increases. With the growth of the total angular momentum, the magnitude of partial energy decreases and converges to a specific value. It was shown that for states with even and odd angular momenta, this quantity has positive and negative values, respectively. Regarding the results, the dipole-dipole interaction revealed different properties from central-like potentials. Since the general characteristics of systems in low dimensions are very different from the bulk systems, the influence of dipole-dipole interaction on the thermodynamic characteristics in metallic nanowires would be an interesting issue for further research.

**Acknowledgements**

We wish to thank Zeinab Rezaei for the useful comments and discussion. The authors also wish to thank the Shiraz University Research Council.

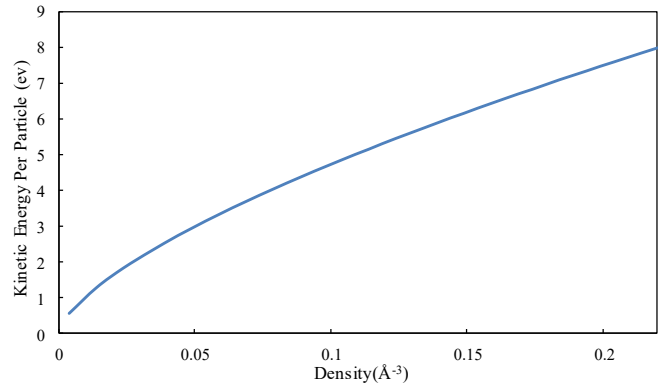

Fig. 1. Kinetic energy per particle as a function of density.

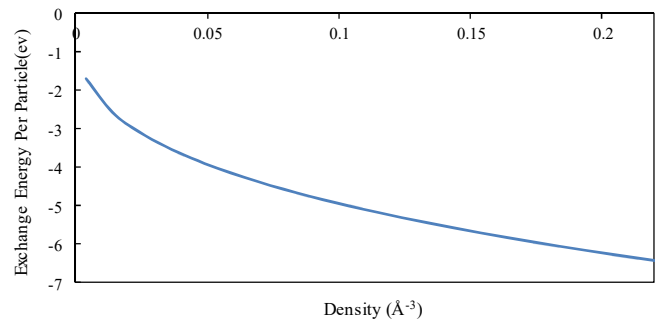

Fig. 2. The exchange energy per particle as a function of density.

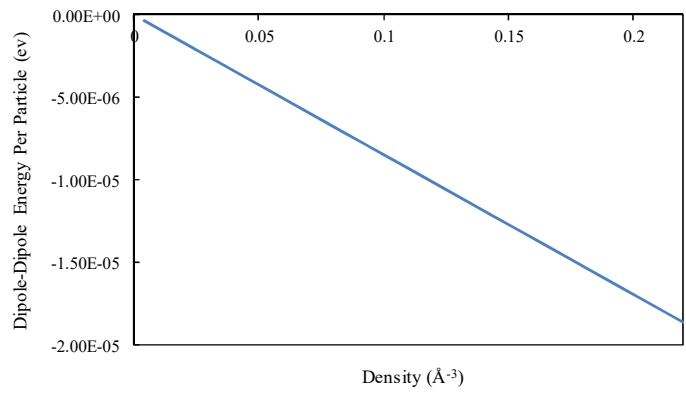

Fig. 3. The energy per particle due to dipole-dipole interaction versus density.

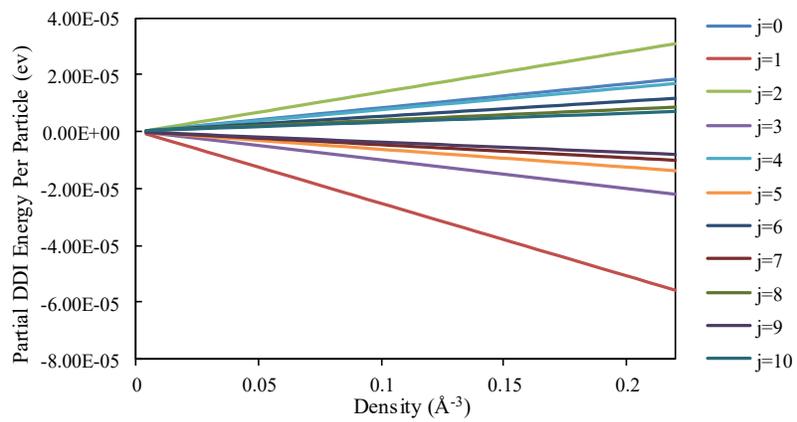

Fig. 4. The contribution of DDI energy per particle for various states ($J=0$ to $J=10$) versus density.

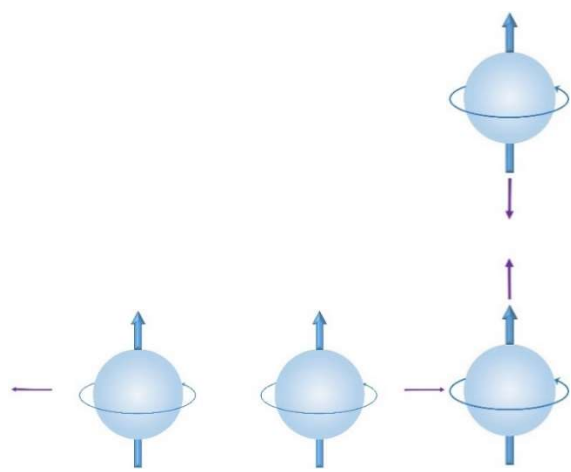

Fig. 5. Dipole-dipole interaction in two different cases.

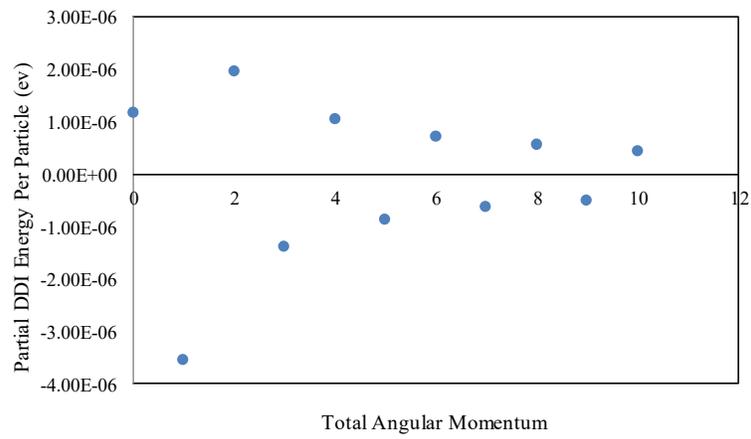

Fig. 6. The contribution of DDI energy per particle for different states in equilibrium density $\rho = 0.014 A^{-3}$.

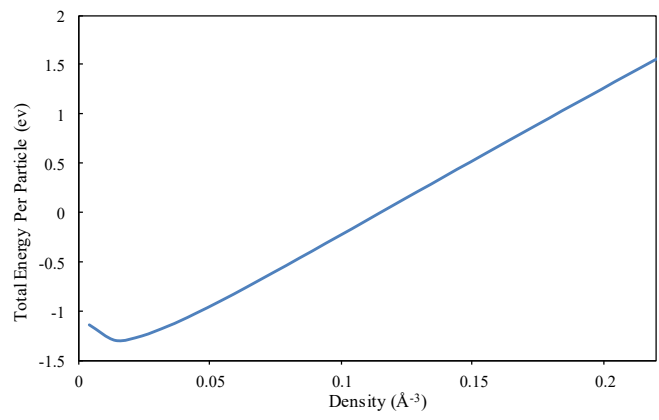

Fig. 7. Density dependence of the ground state total energy per particle.